\documentclass[10pt,journal,compsoc]{IEEEtran}

\usepackage[utf8]{inputenc}
\usepackage[T1]{fontenc}

\usepackage[nocompress]{cite}

\usepackage{graphicx}
\usepackage{url}
\usepackage{multirow}
\usepackage{amsmath}

\usepackage{fancyhdr}

\begin{document}

\newcommand\thetitle{Power Side-Channel Attacks in Negative Capacitance Transistor (NCFET)}
\title{\thetitle}

\author{%
Johann~Knechtel\textsuperscript{*},~\IEEEmembership{Member,~IEEE}, 
Satwik~Patnaik\textsuperscript{*},~\IEEEmembership{Graduate Student Member,~IEEE},
Mohammed~Nabeel\textsuperscript{*},
Mohammed~Ashraf,
Yogesh S.~Chauhan,~\IEEEmembership{Senior Member,~IEEE},
J\"{o}rg~Henkel,~\IEEEmembership{Fellow,~IEEE},
Ozgur~Sinanoglu,~\IEEEmembership{Senior Member,~IEEE}, and
Hussam~Amrouch,~\IEEEmembership{Member,~IEEE}
\IEEEcompsocitemizethanks{%
\IEEEcompsocthanksitem \textsuperscript{\textbf{*}}J.\ Knechtel, S.\ Patnaik, and M.\ Nabeel contributed equally.
\IEEEcompsocthanksitem {J.\ Knechtel, M.\ Nabeel, M.\ Ashraf, and O.\ Sinanoglu are with the Division of Engineering, New York University Abu Dhabi, Saadiyat
Island, 129188, UAE (e-mail: johann@nyu.edu; mtn2@nyu.edu; ma199@nyu.edu; ozgursin@nyu.edu).}
\IEEEcompsocthanksitem {S.\ Patnaik is with the Department of Electrical and Computer Engineering, Tandon School of Engineering, New York University, Brooklyn, NY, 11201, USA (e-mail: sp4012@nyu.edu).}
\IEEEcompsocthanksitem {Y.~S.\ Chauhan is with the Department of Electrical Engineering, IIT Kanpur, Kanpur, India (e-mail: chauhan@iitk.ac.in).} 
\IEEEcompsocthanksitem {J.\ Henkel and H.\ Amrouch are with the Department of Computer Science, Karlsruhe Institute of Technology (KIT), 76131 Karlsruhe, Germany
	(e-mail: henkel@kit.edu; amrouch@kit.edu).}}
}

\markboth{IEEE Micro}%
{Knechtel \MakeLowercase{\textit{et al.}}: \thetitle}

\IEEEtitleabstractindextext{%
\begin{abstract}
Side-channel attacks have empowered bypassing of cryptographic components in circuits. Power side-channel (PSC) attacks have received particular
traction, owing to their non-invasiveness and proven effectiveness. Aside from prior art focused on conventional technologies, this is the first
work to investigate the emerging Negative Capacitance Transistor (NCFET) technology in the context of PSC attacks. We implement a CAD flow for
PSC evaluation at design-time. It leverages industry-standard design tools, while also employing the widely-accepted correlation power analysis
(CPA) attack. Using standard-cell libraries based on the 7nm FinFET technology for NCFET and its counterpart CMOS setup, our
evaluation reveals that NCFET-based circuits are more resilient to the classical CPA attack,
due to the considerable effect of negative capacitance on the switching power.
We also demonstrate that the thicker the ferroelectric layer, the higher the resiliency of the NCFET-based
circuit, which opens new doors for optimization and trade-offs. 
\end{abstract}

}

\maketitle

\renewcommand{\headrulewidth}{0.0pt}
\thispagestyle{fancy}
\lhead{}
\rhead{}
\chead{\copyright~2020 IEEE.
This is the author's version of the work. It is posted here for personal use.
	Not for redistribution.	The definitive Version of Record is published in
IEEE Micro, DOI 10.1109/MM.2020.3005883}
\cfoot{}

\IEEEraisesectionheading{\section{Introduction}\label{sec:introduction}}

\IEEEPARstart{A}{dvances} for CMOS and emerging technologies have led to the ubiquitous use of electronic systems for every aspect of our daily life.
To protect sensitive information within electronic systems, so-called \textit{ciphers} are widely used, which are mathematically unbreakable
cryptographic algorithms.
However, an attacker with physical access
can monitor the interactions of the cipher running on the underlying hardware to make deductions about the secret key.
These physical interactions provide \textit{side-channel information} and related attacks
are known as side-channel attacks (SCAs).
In this work, we focus on the prominent \textit{power side-channel attack (PSC)}, where an attacker analyzes the correlation between power traces
and the cryptographic operations running on the hardware.
Numerous types of PSCs have been demonstrated, including simple power analysis (SPA), differential power analysis
(DPA),
and
correlation power analysis (CPA)~\cite{brier2004correlation}.

\subsection{Negative Capacitance Transistor (NCFET)}
NCFET~\cite{sayeefnano}
is a promising emerging technology.
It overcomes
the fundamental limit in the existing MOSFET technology related to the so-called ``Boltzmann tyranny''~\cite{sayeefnano} that dictates the sub-threshold swing ($SS$) of a transistor to be above $60$mV/dec at room temperature.
This fundamental limit restricts the scalability of the threshold voltage for new nodes, despite the innovations in
transistor structures (e.g.,\ from planar transistors to FinFET, then nanowire, even nanosheet structures). As a result, gains in performance and efficiency with technology scaling become increasingly harder to achieve.

For NCFET,
a thin layer of ferroelectric (FE) material is integrated within the transistor's gate stack. The FE layer behaves as a negative capacitance,
resulting in a voltage amplification instead of a voltage drop as
in conventional transistors. Such higher internal voltage pushes $SS$ to move beyond its fundamental limit.
A point of inflection had occurred when GlobalFoundries
demonstrated that NCFET could be made compatible with the
CMOS fabrication process~\cite{KrivokapicIEDM17}, by utilizing ferroelectricity in $HfO_2$-based materials, which are standard materials for the transistor dielectric.

Importantly, due to the negative capacitance, the total gate capacitance of NCFET is always greater than the conventional/counterpart transistor~\cite{our_ieee_access}.
Hence, at the same operating voltage, NCFET-based circuits consume higher switching power compared to traditional CMOS circuits. 
\textit{This, in turn, necessitates a deeper understanding of security in the context of NCFET technology, in particular concerning PSCs.  
In this work, we are the first to perform such analysis.}

\begin{figure*}[ht]
\centering
\includegraphics[width=0.98\textwidth]{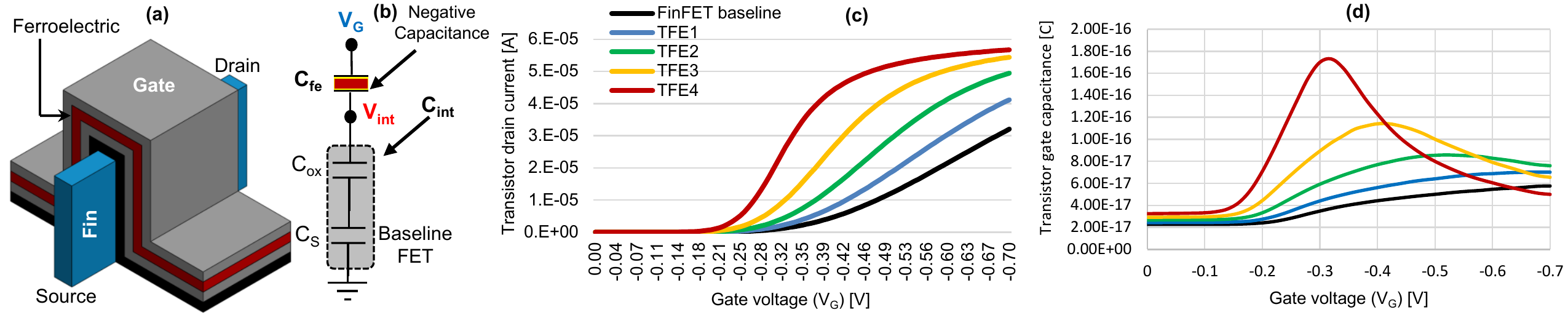}
\caption{(a) NC-FinFET transistor structure, NCFET for short, with ferroelectric layer integrated inside the transistor's gate
stack~\cite{our_ieee_access}. 
(b) Equivalent capacitance series, where the internal voltage exhibits a greater voltage ($V_{internal} > V_G$). 
(c) Drain current across gate voltage for a p-type transistor for four different NCFET cases compared to the FinFET baseline. 
(d) Large increase in the transistor's gate capacitance ($C_{GG}$) due to the negative capacitance effects.}
\label{fig:device_analysis}
\end{figure*}

\subsection{On the Necessity of CAD Flows for Security Assessment of Emerging Technologies}
To assess hardware implementations of ciphers w.r.t.\ PSC attacks, a widely adopted strategy is to implement them using a field-programmable gate array (FPGA).
However, such an FPGA-based evaluation may not capture the resilience of an ASIC implementation accurately, due
to the fundamental differences in the underlying circuit architectures.
For emerging technologies like NCFET, FPGA-based evaluation is not even an option, to begin with,
due to the non-existence of such platforms like NCFET FPGAs.

Hence, CAD flows that allow developers to evaluate the resilience of their cipher implementation for different technologies during
design-time become indispensable.
Such CAD flows
will not only reveal the role that underlying technology plays, but it will also allow for:
(1) The detection of vulnerabilities at design-time, providing an early feedback to improve the cipher implementation;
(2) The accurate analysis of newly-designed cryptographic hardware primitives and/or countermeasures.

\subsection{Related Work}
Only a few works have previously analyzed the resiliency of beyond-CMOS devices against PSC attacks. 
For example, Alasad \textit{et al.}~\cite{alasad2018resilient} studied the resilience of all-spin logic (ASL)-based ciphers.
They showed that the differential power
at the output of hybrid spintronic-CMOS devices, coupled with asymmetric read/write operations in the magnetic tunnel junction (MTJ), poses a security
risk.
For another example, Bi \textit{et al.}~\cite{bi2016tunnel} employed DPA to study tunnel-FET-based implementations
of lightweight ciphers like KATAN32.
    
\subsection{Our Contributions}

This is the \textit{first} work to investigate the resiliency of NCFET against PSC attacks.
It is also the \textit{first} to propose and implement an attack-evaluation flow using commercial tools for emerging technologies.
Prior art relied either on simplified power analysis or SPICE simulations, which are limited to small circuits due to their significant
computational time.

More specifically, our flow enables designers to analyze ciphers (or any other modules), running on emerging technologies, against the powerful CPA attack.
In this work, the resilience of the well-known Advanced Encryption Standard (AES) is evaluated and compared for each NCFET technology setup.
We employ sophisticated NCFET standard-cell libraries and their counterpart baseline library (7nm FinFET) to analyze how the underlying technology
impacts the power profile of the cipher hardware during operation.
By doing so, we succeed in unveiling the role which the thickness of the added ferroelectric layer plays with respect to PSC attacks.
Note that our flow can be easily used for other ciphers as well as for other emerging technologies (i.e., as long as standard-cell libraries
for the technologies are available). 

\section{Impact of NCFET on Power}
\label{sec:NCFET_power}

The FE layer integrated within the transistor gate stack manifests itself as a negative capacitance that provides an internal voltage
amplification inside the transistor ($V_{int} > V_g$). The voltage amplification (Eq.~\ref{eq:voltage_amplification}) is related to both the
internal capacitance of the transistor ($C_{internal}$) and the negative capacitance obtained by the FE layer ($C_{ferro}$).

\begin{align}
\label{eq:voltage_amplification}
& V_{int} = A_V \cdot V_{g} \text { ; } A_V = \frac{|C_{\rm ferro}|}{|C_{\rm ferro}|-C_{\rm internal}} \\
& \text{To ensure no hysteresis: } |C_{\rm ferro}| > C_{\rm internal} \Rightarrow A_V > 1 \nonumber
\end{align} 

The internal voltage amplification magnifies the vertical electric field of the transistor, leading to a higher driving current. As can be
noticed in Fig.~\ref{fig:device_analysis}(c), the drain current of a transistor in NCFET is always higher than the baseline current in the
original CMOS FinFET. In fact, the thicker the FE layer, the higher the drain current.
In this work, we consider four different layer thicknesses of FE,
namely $1$nm, $2$nm, $3$nm, and $4$nm, which we refer to as TFE$1$, TFE$2$, TFE$3$, and TFE$4$, respectively.
The thickness is limited to $4$nm because for higher thicknesses, a hysteresis-free operation in NCFET transistor, which is essential to build logic gates, cannot be ensured any more~\cite{our_ieee_access}. 
The switching power ($P_{switching}$) of any logic is governed by the switching activity ($\alpha$), total capacitance ($C$), operating voltage
($V_{DD}$) and operating frequency ($f$): $P_{switching} \propto  \alpha C V_{DD}^2 f$. Because $C_{\rm ferro}$ exhibits a negative value and
$|C_{\rm fe}| > C_{\rm int}$, the total capacitance of NCFET $C_{NCFET}$ is, in fact, always greater than the FinFET baseline capacitance ($C_{internal}$):  
\begin{align}
\label{eq:cap_NCFET}
C_{NCFET} = \frac{C_{\rm ferro} \times C_{\rm internal} }{C_{\rm ferro} + C_{\rm internal}} > C_{\rm internal}
\end{align}

As Fig.~\ref{fig:device_analysis}(d) shows, NCFET transistors always exhibit a greater gate capacitance ($C_{gg}$). When the FE layer thickness is larger, the gate capacitance becomes higher due to the greater negative capacitance. 
\textit{Therefore, when a circuit is implemented in NCFET technology, it will consume a higher switching power when compared to conventional
	CMOS technology and, thus, its susceptibility to PSC attacks might be different.}

\section{CAD Flow for PSC Evaluation}
\label{sec:CAD_flow}

\begin{figure}[tb]
\centering
\includegraphics[width=\columnwidth]
{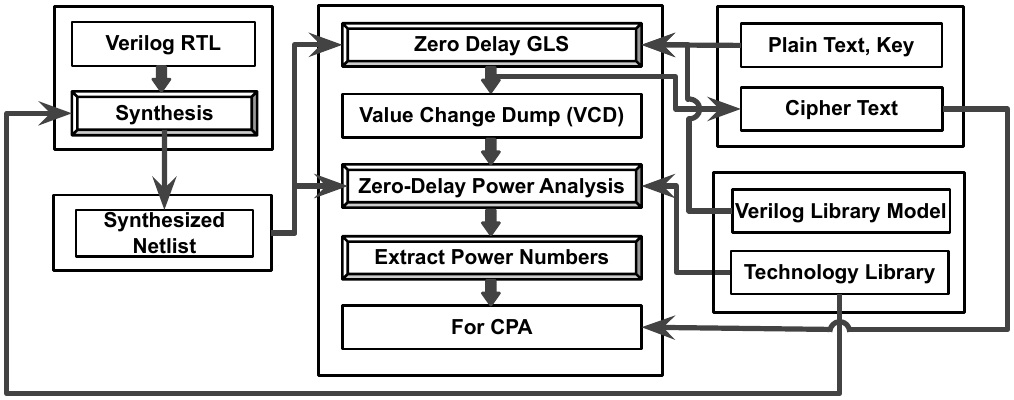}
\caption{CAD flow for PSC evaluation of emerging technologies.}
\label{fig:overview_CAD_flow}
\end{figure}

Next, we present our CAD flow that evaluates cryptographic hardware against PSC attacks
(Fig.~\ref{fig:overview_CAD_flow}).
We focus on AES in this work; however, our flow is not limited to a specific cipher. 
Our flow takes the register-transfer level
(RTL) of the cipher circuit as input, along with the standard-cell library for the technology of interest, and it returns
the minimum number of power traces required to break the cipher, i.e.,~to reveal the secret key.
A greater number of power traces indicates a higher complexity for the attack and, hence, a lower susceptibility of the cipher hardware and/or the technology to PSC attacks. 

\subsection{Simulation-Based Power Analysis}
\label{sec:zero-delay_flow}

First, the cipher RTL is synthesized using the target technology library.
Then, using a testbench, the functionality of the gate-level, post-synthesis netlist
is verified.
After the user specifies a set of plain-texts and a key (or set of keys),
the testbench generates the corresponding set(s) of cipher-texts
required for verification.
During these gate-level simulations,
a Value Change Dump (VCD) file is also generated;
it captures the switching activity of every node within the netlist in a user-defined time resolution (e.g., $1$~ps).
Next, the VCD file is used for power simulation, along with the post-synthesis netlist and library/libraries of interest. 
Note that the \textit{dynamic power} typically dominates the power consumption. This holds even more true for NCFET technology, due to the
important role of negative capacitance for increasing the transistors' capacitance and thus magnifying the switching power; recall
Section~\ref{sec:NCFET_power}.

Instead of full timing simulations,
we leverage zero-delay simulations, for the following reasons.
For zero-delay analysis, all signal transitions occur at the active edge of the clock, where
the peak power values are easier to extract.
For PSC evaluation, in fact, we are mainly interested in the switching power of particular registers; this switching power occurs always at the clock edge,
whereas glitching power occurs only after that.
Thus, ignoring glitching power and related noise
renders the PSC evaluation conservative from a security point of view.
Besides, we consider only the relevant time intervals, i.e., the last round of
AES~\cite{brier2004correlation}.

In short, we obtain the design-time power traces
for AES stepwise
when processing sets of texts for the secret key(s), and we extract the peak-power values for the related last-round operations~\cite{brier2004correlation}.
We have to do so separately for each technology setup considered in this work, i.e.,~for NCFET with different FE configurations and for the
FinFET baseline.

\subsection{Correlation Power Analysis}
\label{sec:our_CPA}

The correlation power analysis (CPA)~\cite{brier2004correlation}
is a powerful attack that uses \textit{Pearson correlation coefficient (PCC)} to measure the relation between predicted and actual power
profiles of a device undergoing some cryptographic operations.
These operations are a function of variable data (i.e., the plain-/cipher-texts to encrypt/decrypt) and the secret key.

\begin{figure}[tb]
\centering
\includegraphics[width=\columnwidth]
{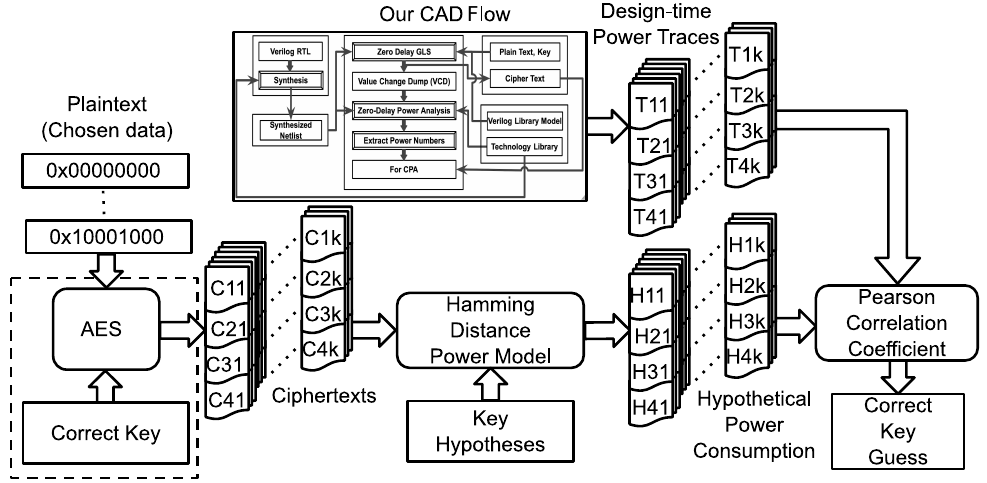}
\caption{Integration of CPA attack in our CAD flow.}
\label{fig:integration_CPA_CAD_flow}
\end{figure}

The integration of the CPA attack in our CAD flow is illustrated in Fig.~\ref{fig:integration_CPA_CAD_flow}.
First, the predicted power consumption is derived from a power model~\cite{brier2004correlation}.
Note that such modeling is to be repeated for all possible candidates of the secret key;
therefore, the resulting values are related to the key hypotheses and are referred to as \textit{hypothetical power values}.
Note that the dynamic power depends largely on switching activities, i.e.,~on rising signal transitions $0 \rightarrow 1$ and falling signal
transitions $1 \rightarrow 0$.
Typically, registers consume a significant portion of the dynamic power during such transitions.
Thus, considering the Hamming distance (HD) for the output of registers before and after switching transitions
is well-established as \textit{HD power model}~\cite{brier2004correlation}.
More specifically, we consider the registers holding the intermediate round texts during the last-round operation to build
up the HD power model.
Importantly, this consideration is also in agreement with the scope of the collected design-time traces.
Note that attacking the first or last round is proven to be both effective~\cite{brier2004correlation}, with
the difference being the use of plain- or cipher-texts as known references required for the HD power model.
In the final step, the design-time power values and 
the hypothetical power values are correlated.
The key candidates are sorted by the PCC, with the candidate exhibiting the highest PCC value
being considered as the correct one.

The above steps are sufficient for an actual CPA attack. For quantifying the susceptibility of the hardware and technology under
consideration, however, these steps have to be conducted throughout multiple trials, e.g., by varying the secret key and the texts,
while tracking the progression of \textit{success rate} (i.e., runs where the secret key is correctly inferred over all
runs); see Sec.~\ref{sec:setup} for further details.

\section{Experimental Investigation}
\label{sec:experiments}

\subsection{Setup for CAD Flow and CPA}
\label{sec:setup}

In our implementation of the CAD flow, we employ commercial tools as follows.
We use \textit{Synopsys VCS M-2017.03-SP1}
for functional simulations at RTL and gate level, \textit{Synopsys DC M-2016.12-SP2}
for logic synthesis, and \textit{Synopsys PrimeTime PX M-2017.06} for power simulations.

For the AES implementation, we leverage a regular RTL, which is working on 128-bit keys and 128-bit texts, using look-up tables for the AES substitution box, and without using any PSC countermeasures.
We provide the AES netlist in~\cite{CPA}.
For the AES circuit implementations using the FinFET baseline technology as well as all NCFET-specific technology setups, we use the same supply voltage $V_{DD} = 0.7V$, 
the same switching activities $\alpha$ (as dictated by \textit{VCS}),
and the same frequency $f = 100 MHz$. Thus, $P_{switching}$ differs only for varying capacitances $C$, given that
$P_{switching} \propto  \alpha C V_{DD}^2 f$.

For the CPA implementation, we extend an open-source C/C++ framework; we provide our version in~\cite{CPA}.
All CPA runs are executed on a high-performance computing (HPC) facility, with 14-core Intel Broadwell processors (Xeon E5-2680) running at 2.4
GHz, and 4 GB RAM are guaranteed (by the Slurm HPC scheduler) for each CPA run.

For the CPA trials, we generate and store ten random keys (128 bits each).  We also generate and store
2,000 random plain-texts (also 128 bits) and, separately for each key, the corresponding 2,000 cipher-texts.  We keep the
sets of keys and corresponding texts the same across the technologies setups; hence, the simulated power values
vary only due to the underlying technology.
Finally, to enable the CPA attack to progress stepwise through the sample space of all power traces (to thoroughly quantify the
success rate depending on the number of traces considered),
we generate and store three batches of permutations for power values and corresponding texts as follows.
Over the course of 1,000 steps, we randomly pick 1,000 sets of power values/texts for each step such that
each step describes 1,000 permutations in multiples-of-two of all 2,000 available pairs of power values/texts.
In total, this results in 1,000,000 permutation sets of increasing size.
For example,
for the first three steps,
we might randomly select the following sets of power values/texts permutations, encoded by indices in the range 1--2,000: \{$734$, $1297$\}, \{$87$, $815$, $562$, $33$\}, and
\{$245$, $734$, $12$, $1395$, $1553$, $94$\}.
As indicated, we independently generate three such batches of 1,000,000 permutation sets, which are employed for three independent CPA trials.
To ensure fair comparisons, these batches are all memorized, stored, and re-applied when conducting the CPA attack for different keys as well as
for different technologies.

\subsection{Setup for NCFET-Specific Library Characterization}
\label{sec:NCFET_lib_characterization}

\begin{figure}[tb]
\centering
\includegraphics[width=0.75\columnwidth]
{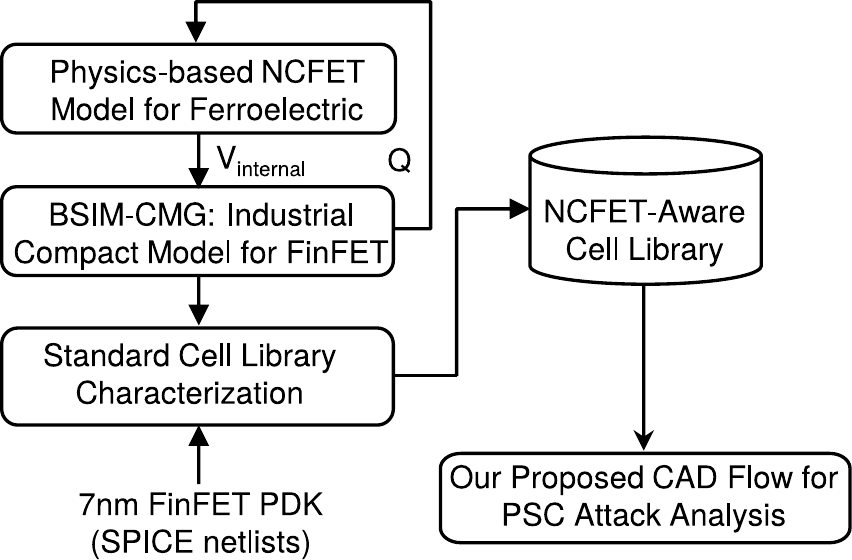}
\caption{NCFET-specific library characterization for 7nm FinFET technology.}
\label{fig:NCFET_flowchart}
\end{figure}

As indicated, a standard-cell library of the target technology is essential.
In this work, we employ the open-source 7nm FinFET cell library as baseline technology~\cite{ClarkMEJ16}.
Then, we characterize the library to create the required NCFET-specific cell libraries~\cite{our_ieee_access} (Fig.~\ref{fig:NCFET_flowchart}).
To achieve that, we use SPICE simulations for post-layout netlists
including parasitic information for a wide range of sequential and combinational standard cells provided within the 7nm PDK~\cite{ClarkMEJ16}.

As is typically done in any commercial standard-cell library, we characterize power and delay of every standard cell under $7$ input signal slews and $7$ output
load capacitances. 
To model the impact of FE layer and the negative capacitance effect on the electrical characteristics of nMOS and pMOS transistors, we employ the
state-of-the-art physics-based NCFET model~\cite{pahwa2016analysis}. We integrate the model in a self-consistent manner within BSIM-CMG, which is the industry standard compact model of FinFET technology~\cite{our_ieee_access}.
The material properties, required for the NCFET physics-based compact model, were obtained from the experimental measurement presented in~\cite{MullerNanoLett12}.

For a comprehensive analysis, recall that we consider four different cases for FE thicknesses, ranging from $1$nm to $4$nm and referred to as
TFE$1$, TFE$2$, TFE$3$ and TFE$4$, respectively.
Our cell libraries are fully compatible with commercial CAD tools.
Therefore, we can directly deploy them within
our flow
to investigate the resiliency of AES (or other ciphers) against PSC attacks when NCFET technology is used in comparison to the FinFET baseline technology.

\begin{table*}[tb]
\centering
\scriptsize
\setlength{\tabcolsep}{0.8mm}
\caption{Statistics on Traces Required on Average for Successful CPA Runs}
\label{tab:results}
\begin{tabular}{*{19}{c}}
\hline
\multirow{2}{*}{\textbf{Technology}}
& \multicolumn{6}{c}{\textbf{50\% Success Rate}}
& \multicolumn{6}{c}{\textbf{90\% Success Rate}}
& \multicolumn{6}{c}{\textbf{99.9\% Success Rate}}
\\
\cline{2-19}
\multirow{2}{*}{\textbf{Setup}}
& \multicolumn{2}{c}{\textbf{Trial 1}}
& \multicolumn{2}{c}{\textbf{Trial 2}}
& \multicolumn{2}{c}{\textbf{Trial 3}}
& \multicolumn{2}{c}{\textbf{Trial 1}}
& \multicolumn{2}{c}{\textbf{Trial 2}}
& \multicolumn{2}{c}{\textbf{Trial 3}}
& \multicolumn{2}{c}{\textbf{Trial 1}}
& \multicolumn{2}{c}{\textbf{Trial 2}}
& \multicolumn{2}{c}{\textbf{Trial 3}}
\\
\cline{2-19}
& Avg & Std Dev
& Avg & Std Dev
& Avg & Std Dev
& Avg & Std Dev
& Avg & Std Dev
& Avg & Std Dev
& Avg & Std Dev
& Avg & Std Dev
& Avg & Std Dev
\\
\hline
FinFET Baseline
& 501.5 & 46.8 
& 502.0 & 46.6 
& 502.2 & 45.7 
& 693.0 & 90.3 
& 688.8 & 80.6 
& 693.6 & 86.4 
& 985.2 & 156.9 
& 991.4 & 148.9 
& 983.0 & 151.8 
\\ \hline
TFE$1$
& 503.4 & 47.2 
& 505.6 & 48.3 
& 504.3 & 47.3 
& 698.4 & 93.5 
& 696.2 & 92.6 
& 695.2 & 88.1 
& 987.6 & 156.2
& 988.0 & 144.3
& 982.6 & 151.9
\\ \hline
TFE$2$
& 509.5 & 49.7 
& 509.6 & 52.5 
& 507.8 & 46.2 
& 706.8 & 94.8 
& 702.2 & 94.7 
& 699.6 & 87.9 
& 995.2 & 151.5 
& 999.2 & 159.9 
& 985.0 & 158.7
\\ \hline
TFE$3$
& 519.6 & 52.2 
& 520.4 & 48.3 
& 519.6 & 51.4 
& 724.4 & 107.3 
& 718.8 & 101.2 
& 721.6 & 103.8 
& 1,011.6 & 156.4 
& 1,026.6 & 164.7 
& 1,013.4 & 165.9
\\ \hline
TFE$4$
& 547.8 & 58.9 
& 546.0 & 57.3 
& 545.6 & 56.2 
& 759.2 & 117.4 
& 758.6 & 111.8 
& 760.8 & 122.1 
& 1,066.0 & 192.3 
& 1,065.8 & 185.9 
& 1,068.4 & 187.8
\\ \hline
\end{tabular}
\\[1mm]
For a fair comparison, the same selection of traces is considered across all technology setups.
Results are obtained considering all ten random-but-reproducible keys and for three independent trials employing our scheme of random-but-reproducible
permutation of traces.
\end{table*}

\subsection{Results} 
\label{sec:results_NCFETs}

The primary goal of this study is to investigate the resilience of the NCFET technology against PSC attacks, i.e.,~the classical and powerful CPA attack in particular. 
The key finding of this study is the following: the thicker the FE layer,
the more resilient becomes the device.
This is because for thicker FE layers, the negative capacitance effects become greater and, thus,
the dynamic power becomes more dominant and more varied, rendering the classical CPA more difficult.
Next, we discuss our findings in more detail.

\begin{figure}[tb]
\centering
\includegraphics[width=\columnwidth]{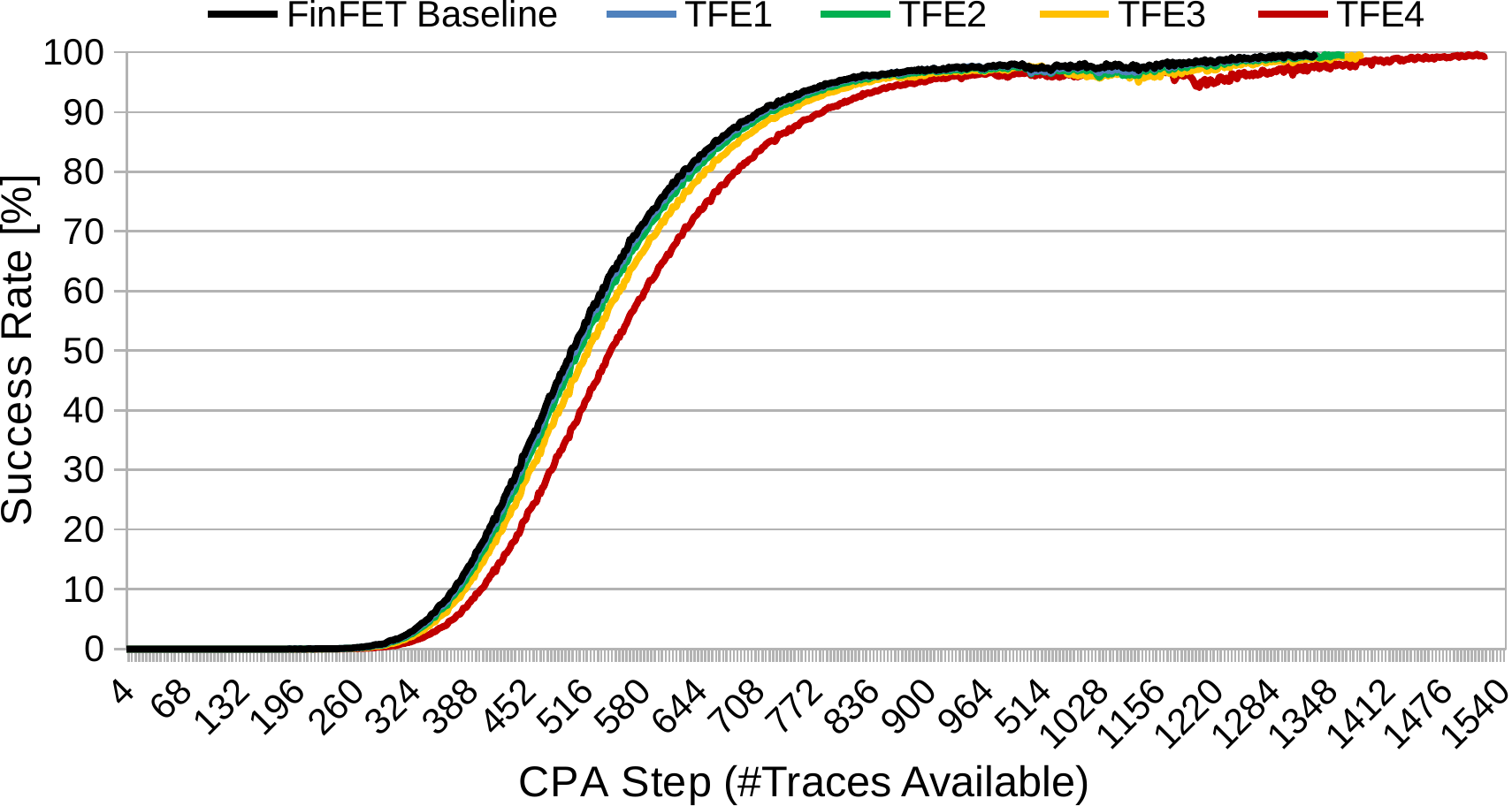}
\caption{CPA progression in terms of success rates over traces being available for consideration.
Success rates are averaged across all ten random-but-reproducible keys and across three independent trials
considering the random-but-reproducible permutation sets for power values and cipher-texts.
Thus, there are 30,000 CPA runs underlying \textit{for each data point} for all curves.
This ensures a fair and robust comparison across technology setups, along with thorough sampling across traces being available for
consideration in the CPA framework.}
\label{fig:CPA_progress}
\end{figure}

In Table~\ref{tab:results}, we disclose the number of traces required for the CPA attack to infer the correct key for varying success rates
(e.g., for a $50\%$ success rate, $500$ out of $1,000$ runs provide the correct key).
We report the averages and the variations (i.e., standard deviations) across all ten random-but-reproducible keys,
and we do so for three independent trials employing our scheme of random-but-reproducible permutations for power values/cipher-texts.
In Fig.~\ref{fig:CPA_progress}, we illustrate the progression of success rates over CPA steps (i.e., traces considered),
again averaged over the same ten keys, and further averaged over the same three trials.
Therefore, we enable a truly fair and robust comparison across all technology configurations.

We conclude from both Table~\ref{tab:results} and Fig.~\ref{fig:CPA_progress} that TFE$4$ represents the most resilient setup and the FinFET baseline the weakest setup, respectively.
More specifically, TFE$4$ is on average 8.13\% more resilient than the FinFET baseline (i.e., for a success rate of 99.9\%).
We further conclude that the increase of both the average resilience and the variation of resilience from FinFET baseline to TFE$4$ is not linear, but more pronounced toward TFE$4$, which is explained next.

\begin{figure}[tb]
\centering
\includegraphics[width=\columnwidth]{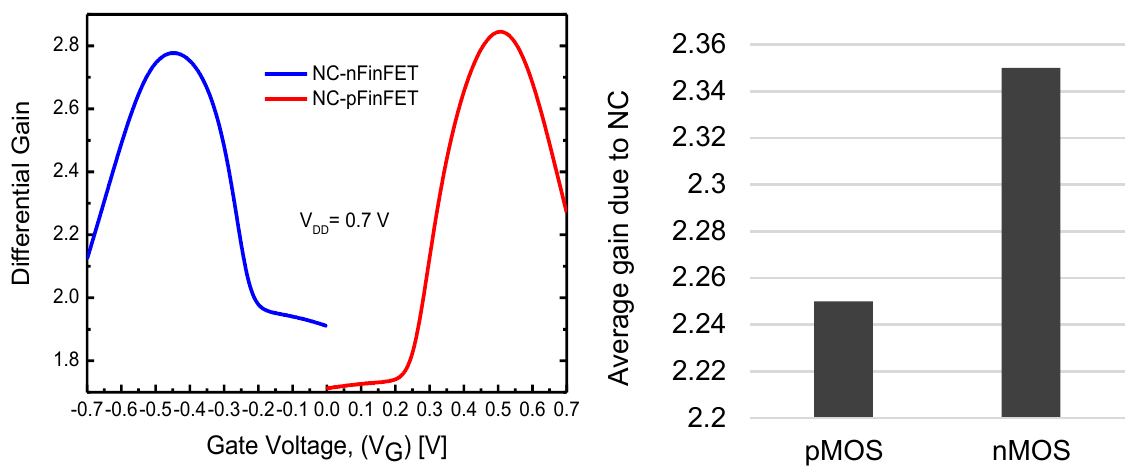}
\caption{Differential gain due to negative capacitance in nMOS and pMOS FinFET transistors (left) and the corresponding average gain (right).}
\label{fig:device_gain}
\end{figure}

As demonstrated in Fig.~\ref{fig:device_analysis}(d), a larger thickness of the FE layer results in a higher, non-linear increase in the transistor gate capacitance; the most significant increase occurs for TFE$4$.
As explained earlier as well, the dynamic power depends largely on the capacitances across the whole circuit.
In fact,
the total capacitances (of standard cells) and their impact on dynamic
power become both more dominant and more varied for thicker FE layers.
More specifically, despite the FE layer being identical for the nMOS and pMOS transistors in all the standard cells, the related negative capacitance
	effects play out differently.
	Essentially, this is because of different capacitance matching between the FE layer
	added to the transistor gate and the MOS capacitance of the underlying transistor
	and, hence, varying differential gains in nMOS and pMOS
	transistors.
The differential gain ($A_V$) and the resulting average gain ($A_{avg}$) are given in Eq.~\ref{eq:voltage_amplification_nmos_pmos} and
related simulation results shown in Fig.~\ref{fig:device_gain}.
As can be observed, nMOS transistors exhibits a higher average gain than pMOS transistors.
\begin{align}
\label{eq:voltage_amplification_nmos_pmos}
&A_V = \frac{\partial{V_{int}}}{ \partial{V_G}}  \text{ , where } A_{avg} = \frac{1}{V_G}\int_{0}^{V_G} A_V~dV_G
\end{align}

Such dis-proportionality leads, in turn, to a varying impact on the switching power for nMOS and pMOS transistors.
Thus, we observe that the switching power for rise versus fall transitions of the AES round-operation registers varies to a greater extent
for TFE$4$ when compared with the FinFET baseline (Table~\ref{tab:dynamic_power}).
The varying impact which the negative capacitance plays for pMOS
versus nMOS transistors, especially for thicker FE
layers, thus represents the root-cause for the higher resilience observed with NCFET technology---the greater the variations for the rise versus fall
switching power are, the more ``noisy'' the power samples become, and the lower the accuracy becomes for the classical CPA based on the well-established HD power model.

\begin{table}[tb]
\scriptsize
\caption{Power Values for AES Round-Operations Registers}
\label{tab:dynamic_power}
\centering
\begin{tabular}{cccc}
\hline
\multicolumn{2}{c}{\textbf{Power Components}} & \textbf{FinFET Baseline} & \textbf{TFE$4$} \\ \hline
\multicolumn{2}{c}{Static or Leakage} & 3.45E-10 & 3.17E-10 \\ \hline
\multirow{3}{*}{$0 \rightarrow 1$ Transitions} & Clk Rise & 4.89E-07 & 6.94E-07 \\ \cline{2-4} 
& Switching & 1.70E-06 & 3.89E-06 \\ \cline{2-4}
& Total & 2.19E-06 & 4.59E-06 \\ \hline
\multirow{3}{*}{$1 \rightarrow 0$ Transitions} & Clk Rise & 4.89E-07 & 6.94E-07 \\ \cline{2-4} 
& Switching & 1.55E-06 & 3.17E-06 \\ \cline{2-4}
& Total & 2.04E-06 & 3.86E-06 \\ \hline
{$1 \rightarrow 1$, $0 \rightarrow 0$} & Clk Rise & 4.89E-07 & 6.94E-07 \\ \cline{2-4} 
Transitions & Total & 4.89E-07 & 6.94E-07 \\ \hline
\end{tabular}
\end{table}

\subsection{Discussion}

In Fig.~\ref{fig:CPA_progress}, note that success rates are dropping momentarily toward the end of the curves.
This is because the longer the CPA attack progresses, the fewer
keys remain unresolved,
and those remaining ``resilient keys'' tend to incur lower success rates over
longer periods, resulting in these momentary drops.
Such varying resilience across keys is also observable from the standard deviations reported in
Table~\ref{tab:results}.
We caution that
this finding should \textit{not} motivate to favor specific, seemingly ``resilient keys'' for actual applications---such outcomes are highly dependent
on the architecture, RTL, and gate-level implementation of the cipher, as well as the selection of plain-/cipher-texts.
Even for very same sets of texts, which we have used here for all experiments (to enable a fair comparison), we naturally expect and have indeed observed varying
distributions for the bit-level flips across the last-round texts of the AES cipher when using different keys.
These varying distributions, among other aspects, can impact the accuracy of the CPA.
We note that some studies are investigating the role of keys and texts thoroughly, e.g., based on collision, confusion, and/or
statistical modeling~\cite{fei15}, but we also argue that further consideration of such analytical aspects is out of scope for this
manuscript.
In any case,
our findings on NCFET are not contradicted; this is because our findings are centered on average results obtained across all ten keys considered
as well as across three independent CPA trials.

As indicated before, our flow is applicable to any technology, be it CMOS, NCFET, or other emerging devices, as long as those devices are
supported by commercial CAD flows, especially by means of standard-cell libraries being available.
One important aspect for different technologies are their different
operating frequencies.
Recall that CPA considers only the peak power for relevant registers' transitions; therefore, while different operating frequencies will scale
the power amplitude/values, it will not hinder correlation within the CPA framework.
In fact, we have conducted a selection of the same experiments underlying Table~\ref{tab:results} for various frequencies, and we observed the
same number of traces for all configurations.

While the absolute numbers of traces required and their differences across setups may seem small
to some readers (e.g., those familiar with PSC attacks on CMOS devices in the field), the related findings are significant nevertheless.
As they are grounded on physics-based compact models and thoroughly-characterized standard-cell libraries,
along with zero-delay power simulation using commercial-grade tools, these  ``ideal case'' findings represent \textit{firm boundaries} for future
CPA attack on NCFET devices in the field.
That is, given the inevitably noisy behaviour of devices in the field, an attacker cannot perform any
better than what we note here (i.e., assuming the same operating conditions).
Thus,
from a security-enforcing designer's perspective, these findings 
can serve well as conservative guidelines, e.g., for security schemes like dynamic key updating~\cite{taha2014key}.

As indicated, the root-cause for the observed resilience of NCFET technology against the CPA attack are the variations for rise versus fall
switching power being more pronounced for thicker FE layers.  While this root-cause and the resulting effect of power samples becoming more ``noisy'' 
would have a detrimental impact on other PSC attacks as well, a related quantitative study is scope for future work.

\section{Conclusion}
\label{sec:conclusion}

In this work, for the first time, we have investigated the resiliency of NCFET technology against power side-channel attacks.
To achieve that, we
have implemented a CAD flow that enables designers to analyze
the complexity for extracting the secret key from their
cryptographic hardware of choice (AES in our case).
To properly capture the impact that NCFET technology has on power traces, NCFET-specific libraries were created based on accurate
physics-based models. 
Our analysis reveals that NCFET technology renders AES more resilient against classical correlation power
attacks, and the thickness of the ferroelectric layer inside the NCFET transistor plays a major role:
the thicker the layer, the higher the resiliency, due to the greater effects of the negative capacitance on rise/fall switching power.

\section*{Acknowledgments}
\label{sec:ack}

This work was supported in part by the Center for Cyber Security (CCS) at New York University Abu Dhabi (NYUAD).  The work of Satwik
Patnaik was supported by the Global Ph.D.\ Fellowship at NYU/NYUAD.
Besides, parts of this work were carried out on the HPC facility at NYUAD.

\begin{IEEEbiographynophoto}{Johann Knechtel}
received the M.Sc.\ degree in information systems engineering (Dipl.-Ing.) and the Ph.D.\ degree in computer engineering (Dr.-Ing., summa cum laude) from TU Dresden, Germany, in 2010 and 2014, respectively. 
He is a Research Scientist with New York University Abu Dhabi, United Arab Emirates. 
His research interests cover VLSI physical design automation, with particular focus on emerging technologies and hardware security.
\end{IEEEbiographynophoto}

\begin{IEEEbiographynophoto}{Satwik~Patnaik}
received the B.E.\ degree in electronics and telecommunications from the University of Pune, India, and
the M.Tech.\ degree in computer science and engineering with a specialization in VLSI design from the Indian Institute of Information Technology and Management, Gwalior, India. 
He is currently pursuing the Ph.D.\ degree
with the Department of Electrical and Computer
Engineering, Tandon School of Engineering, New
York University, Brooklyn, NY, USA. 
His current research interests include
hardware security, trust and reliability issues for CMOS and emerging devices
with particular focus on low-power VLSI Design.
Mr. Patnaik received the Bronze Medal in the Graduate Category at the ACM/SIGDA Student Research Competition held at ICCAD 2018. 
\end{IEEEbiographynophoto}

\begin{IEEEbiographynophoto}{Mohammed~Nabeel}
received his Bachelors degree in electrical and electronics engineering from National Institute of
Technology--Calicut, India.
Mr.\ Nabeel is currently working as a Research Engineer at Center for Cyber Security at New York University Abu Dhabi (CCS-NYUAD). Apart from
working on research in the field of hardware security, he also focuses on implementing and prototyping the research ideas in Chip. He has around
12 years of industry experience in chip design -- specialized in Micro architecture, protocol know-how, RTL design, Synthesis, Static Timing
Analysis and post silicon bring up. Prior to joining CCS-NYUAD, he worked at Texas Instruments, where he worked on chips targeted for IoT and
Automotive and prior to that was with Qualcomm, where he worked on chips targeted for mobile phones and data cards.
\end{IEEEbiographynophoto}

\begin{IEEEbiographynophoto}{Mohammed~Ashraf}
received the bachelor’s degree
in electronics and telecommunication engineering from the College of Engineering Trivandrum,
Thiruvananthapuram, India, in 2005.
He is a Senior Physical Design Engineer from
India. 
He carries an experience of ten years in
the VLSI industry. 
He has worked with various
multinational companies like NVIDIA Graphics,
Santa Clara, CA, USA, Advanced Micro Devices,
Santa Clara, and Wipro Technologies, Bengaluru,
India. 
He worked also with Dubai Circuit Design,
Dubai Silicon Oasis, Dubai, United Arab Emirates. 
He is currently a Research Engineer with
the Center for Cyber Security, New York University Abu Dhabi, United Arab Emirates.
\end{IEEEbiographynophoto}

\begin{IEEEbiographynophoto}{Yogesh S.~Chauhan}
(SM'12) is an associate professor at Indian Institute of Technology Kanpur (IITK), India. He was with Semiconductor Research \& Development Center at IBM Bangalore during 2007 – 2010; Tokyo Institute of Technology in 2010; University of California Berkeley during 2010-2012; and ST Microelectronics during 2003-2004. His group is also involved in developing compact models for FinFET, Nanosheet/Gate-All-Around FET, FDSOI transistors, Negative Capacitance FETs and 2D FETs. He is the Editor of IEEE Transactions on Electron Devices and Distinguished Lecturer of the IEEE Electron Devices Society. He is the member of IEEE-EDS Compact Modeling Committee and fellow of Indian National Young Academy of Science.
\end{IEEEbiographynophoto}

\begin{IEEEbiographynophoto}{J\"{o}rg~Henkel}
 (M'95-SM'01-F'15) is the Chair Professor for Embedded Systems at Karlsruhe Institute of Technology. Before that he was a research staff member at NEC Laboratories in Princeton, NJ. He received his diploma and Ph.D. (Summa cum laude) from the Technical University of Braunschweig. His research work is focused on co-design for embedded hardware/software systems with respect to power, thermal and reliability aspects. He has received six best paper awards throughout his career from, among others, ICCAD, ESWeek and DATE. For two consecutive terms he served as the Editor-in-Chief for the ACM TECS. He is a Fellow of the IEEE.
 \end{IEEEbiographynophoto}

\begin{IEEEbiographynophoto}{Ozgur~Sinanoglu}
received the first B.S.\ degree in electrical and electronics engineering and the second B.S.\ degree in computer engineering from Boğaziçi University, Istanbul, Turkey, in 1999, and the M.S.\ and Ph.D.\ degrees in computer science and engineering from the University of California at San Diego, CA, USA, in 2001 and 2004, respectively.  
He is a Professor of electrical and computer engineering with New York University Abu Dhabi (NYU Abu Dhabi), United Arab Emirates. 
He has industry experience with TI, Dallas, TX, USA, IBM, Armonk, NY, USA, and Qualcomm, San Diego, CA, USA. 
His recent research in hardware security and trust is being funded by U.S. National Science Foundation, U.S. Department of Defense, Semiconductor Research Corporation, Intel Corp, and Mubadala Technology. 
His research interests include design-for-test, design-for-security, and design-for-trust for VLSI circuits, where he has more than 180 conference and journal papers, and 20 issued and pending U.S. Patents. 
\end{IEEEbiographynophoto}

\begin{IEEEbiographynophoto}{Hussam~Amrouch}
(S'11, M'15) is a Research Group Leader at the Karlsruhe Institute of Technology, Germany. He is leading the Dependable Hardware research group. He received his Ph.D. degree (Summa cum laude) from KIT in 2015. His main research interests are emerging technologies, design for reliability from physics to systems and machine learning. He holds seven HiPEAC Paper Awards. He currently serves as Associate Editor at Integration, the VLSI Journal. He has served in the technical program committees of many CAD conferences like DAC, ICCAD, etc. and as a reviewer in many journals like T-ED, TCAS-I, TCAS-II, TC, etc. He has around 80 publications in multidisciplinary research areas across the computing stack; semiconductor physics, computer-aided design, computer architecture, and circuit design.

\end{IEEEbiographynophoto}

\end{document}